\documentclass{appolb}
\usepackage{graphicx}
\usepackage{amsmath,bm}
\usepackage{xcolor}
\usepackage[colorlinks,pdfstartview=FitH]{hyperref}
  \definecolor{blue2}{rgb}{0.0, 0.2, 0.4}
  \hypersetup{linkcolor=blue2,citecolor=blue2,filecolor=black,urlcolor=blue2}
\usepackage{url}
\usepackage{bookmark} 
\usepackage{cite}
\usepackage{custom}

\begin{document}
\title{QCD critical point and
  hydrodynamic fluctuations
  in relativistic fluids%
\thanks{Presented at the 63rd Cracow School of Theoretical Physics,
  Zakopane, Poland, September 17-23, 2023}%
}
\author{Mikhail Stephanov
\address{Department of Physics and Laboratory for Quantum Theory at the Extremes, University of Illinois, Chicago, IL 60607}
\address{Kadanoff  Center  for  Theoretical  Physics,  University  of  Chicago,  Chicago,  IL  60637}
}
\maketitle
\begin{abstract}

  These lecture notes consist of two major connected parts. The first
  part (Sections 1, 2), after a brief historical introduction, deals
  with the physics of critical points in thermodynamic {\em
    equilibrium}. The features of the fluctuations relevant for the
  QCD critical point search are highlighted. The second part (Sections
  3, 4) focuses on the recent developments in the description of the
  fluctuation {\em dynamics} especially relevant for the QCD critical
  point search in heavy-ion collisions.

\end{abstract}


\section{Introduction}

\begin{figure}[h]
  \centering
  \includegraphics[width=15em]{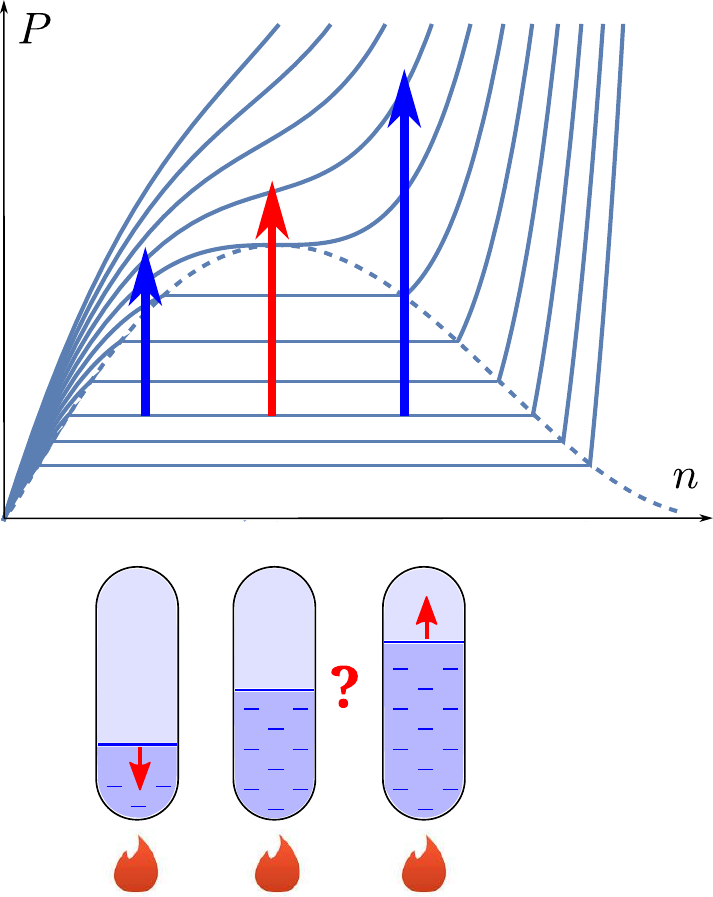}
  \caption{A schematic representation of Cagniard de la Tour's
    experiment, with the heating ``histories'' indicated by the
    vertical (fixed volume) lines on the pressure-density diagram
    (solid lines are isotherms). One can think of this experiment as a
    (density) scan of the phase diagram. On the left: if a small amount of
    fluid is placed in a sealed container and heated the fluid
    evaporates and the meniscus goes down.  On the right: if the fluid
    fills most of the container the fluid expands and the meniscus
    goes up.  There must be a certain intermediate, critical filling
    fraction at which evaporation and expansion compete. What happens
    to the meniscus at this point?}
  \label{fig:vdw-lg-scan}
\end{figure}

Almost exactly 200 years ago Cagniard de la Tour performed an
experiment which led to the discovery of critical points in a number
of liquids~\cite{delaTour:1823}.  One can think about this groundbreaking
experiment as an attempt to address the following question: What
happens if a liquid is heated in a sealed container to such
(temperature and pressure) conditions that
its propensity to evaporate competes with
its propensity to expand, as illustrated in Fig.~\ref{fig:vdw-lg-scan}.

Cagniard de la Tour discovered that the boundary (meniscus) separating
liquid and gas phases disappears. This phenomenon occurred
for all liquids he experimented with, including water. He also
described the phenomenon now called {\em critical opalescence}.

Michael Faraday, at the time, was working on the
problem of liquefying gases --- his report on the liquefaction of
chlorine appears in the same issue with the English translation of
Cagniard de la Tour's article \cite{delaTour:1823}.  Faraday took
great interest in Cagniard de la Tour's discovery, in particular, in
its intriguing implication of the continuity between liquid and
gaseous phases. Curiously, Faraday was frustrated that Cagniard de la
Tour had not named the novel phenomenon and
struggled to come up with a name he needed to refer to a ``point at which the
fluid \& its vapour become one according to a law of continuity''
\cite{Goudaroulis:1994}.


Full understanding of the ubiquity of the phenomenon and its
logical consequence -- continuity between ``liquid and gaseous states'' was
solidified after systematic quantitative studies
by Thomas Andrews~\cite{Andrews:1869}, who also coined
the name ``critical point'' we use today.

The desire to explain the liquid-gas continuity and the critical point
led Leiden University Ph.D. student van der Waals \cite{vdWaals:1873}
to discover a model of the equation of state based on, revolutionary
at the time, molecular description of matter. The law of corresponding
states formulated by van der Waals to describe near-critical fluids
paved the way to the concept of universality of critical phenomena.

The phenomenon of critical opalescence 
was understood by Smoluchowski \cite{Smoluchowski:1908} as an effect
of density fluctuations. This discovery was followed by quantitative
theoretical description by Einstein \cite{Einstein1910}.

The ubiquity of critical phenomena led Landau to develop the
well-known classical theory of phase transitions
\cite{Landau:1937obd}.
But it took much
more work to understand the crucial role of fluctuations. This work
includes the pioneering contributions by Fisher, Kadanoff, and Wilson,
which lead to the modern understanding of the (fluctuating) field
theory (including quantum field theory) in terms of such essential
concepts as scaling and renormalization group.

Thermodynamic critical points are ubiquitous in Nature. Not only
practically all fluids possess such a point, but similar
phenomena occur in completely different physical systems, such as, e.g.,
ferromagnets. The universality of critical phenomena extends over a vast
range of physically different systems.

\begin{figure}[th]
  \centering
  \includegraphics[height=15em]{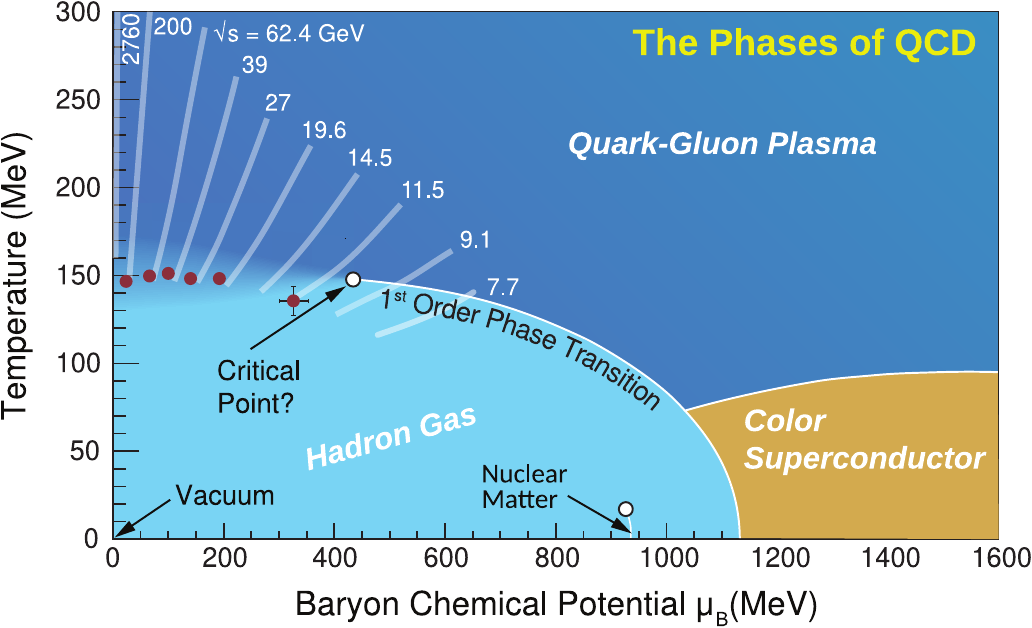}
  \caption{A schematic rendering of the QCD phase diagram as it is
    currently understood or conjectured~\cite{Bzdak:2019pkr}.
    Superimposed are expansion
    ``histories'' of fireballs created in heavy-ion collisions at
    varying energies used in the bean-energy scan experiments (white
    lines labeled by $\snn$ in GeV). The experimental freezeout points
    (red) are illustrating the dependence of the freezeout point on
    $\snn$.}
  \label{fig:phase-diagram}
\end{figure}

Hot and dense QCD matter dominating the Universe moments after the Big
Bang, or recreated in heavy-ion collision experiments, is a fluid.
It is natural to ask if this fluid also possesses
a critical point. This is a non-trivial question first of all because
this fluid, unlike the fluids in which critical points have been
observed so far, is {\em relativistic}. In that respect, we are
especially interested in the critical point which is related to
deconfinement and chiral restoration transition, and we call this
point QCD critical point.\footnote{The critical point of nuclear
  matter (see Fig.~\ref{fig:phase-diagram}) is also of
  interest. However, its existence is not in question and its location
  at temperature of order 10 MeV (commensurate with the binding energy
  of nuclear matter), much smaller than nucleon rest energy, makes it
  somewhat similar to non-relativistic liquid-gas critical
  points.}

There are two necessary prerequisites for the existence of the
critical point. One of them is the existence of two phases (e.g.,
liquid and gas). QCD matter does, indeed, possess such two phases, shown in
Fig.~\ref{fig:phase-diagram} -- Hadron
Resonance Gas (HRG) and Quark Gluon Plasma (QGP), the latter
having liquid properties~\cite{ZAJC2008283c}.
Another prerequisite is the continuity between these two
phases. Such a continuity has been, indeed, by now rather firmly
established by first-principle lattice calculations at zero net baryon
density, or $\mu_B=0$ \cite{Aoki:2006we}.

Does the crossover separating the two QCD phases at zero $\mu_B$
turn into a first order transition at some finite value of $\mu_B$,
and what is that value? I.e., where is the critical
point on the QCD phase diagram (see
Fig.~\ref{fig:phase-diagram})? Unfortunately, this question cannot, at
this time, be answered by a lattice calculation due to the notorious
sign problem.

However, this question can be addressed experimentally, by scanning
QCD phase diagram in heavy-ion collision experiments \cite{STAR:2010vob}, as illustrated
in Fig.~\ref{fig:phase-diagram}.
The sought signatures of the
critical point are based on the {\em non-monotonic} dependence of
fluctuations on the collision energy, $\snn$
\cite{Stephanov:1998dy,Stephanov:2004wx,Bzdak:2019pkr}.

While fluctuations in
thermodynamic {\em equilibrium} near critical points, and their role
in such critical phenomena as critical opalescence have been long and
well understood, heavy-ion collisions present an additional challenge --
fluctuation dynamics.

QCD matter (QGP) fireball created in heavy-ion collisions evolves
sufficiently slowly that it can be described by hydrodynamics. However, the
separation of scales is not astronomical, as in usual condensed matter
contexts. As a result, fluctuations, which take time to relax to
evolving equilibrium during the expansion, could deviate
substantially from equilibrium values at freezeout.

After a brief introduction to equilibrium fluctuations (which can be
also found elsewhere, e.g.,\cite{Bzdak:2019pkr})
these lectures will describe the recent progress achieved in the
dynamical description of fluctuations.



\section{Fluctuations in thermodynamics, critical fluctuations}

\subsection{Relation between equation of state and
  fluctuations}

In thermodynamic equilibrium (by definition) the probability of the
system to be found in a given {\em microscopic}, i.e., quantum, state
depends only on the conserved quantum numbers of this state, such as
energy, momentum, charge, etc. Therefore, the probability of a given
{\em macroscopic} state, characterized by a given set of macroscopic
variables, such as energy, is proportional to the number of
microscopic states with these values of the macroscopic variables.
That probability is equal to the exponential of the entropy, by
definition. Therefore, the dependence of the entropy on energy, etc,
i.e., the equation of state (EOS), determines the probability of the
fluctuations for a system in thermodynamic equilibrium
\cite{landau2013statistical}.

This fundamental relationship between fluctuations and EOS is at the
core of Einstein's description~\cite{Einstein1910} of the density
fluctuations, whose singular behavior near critical points
explains the phenomenon of critical opalescence~\cite{Smoluchowski:1908}.

The entropy in question is the entropy of the open system
which can exchange conserved quantities, such as energy and
charge with the surroundings (thermodynamic bath) at temperature $T$
and chemical potential $\mu$. The corresponding probability
distribution for the values of energy density $\epsilon$ and charge
density $n$ is given by
\begin{equation}
  \label{eq:P=eS}
  \mathcal P(\epsilon,n) \sim \exp\left\{
   V[s(\epsilon,n)-\beta\epsilon + \alpha n]   
      \right\}
\end{equation}
where $\beta=1/T$, $\alpha=\mu/T$ and $V$ is the volume of the
thermodynamic system. In the thermodynamic limit,
i.e., for large $V$, the probability is sharply peaked around the
maximum determined by the familiar conditions:
\begin{equation}
  \label{eq:ds/de=b}
  \left(\frac{\partial s}{\partial\epsilon}\right)_{\!n}-\beta=0,
  \quad
  \left(\frac{\partial s}{\partial n}\right)_{\!\epsilon}+\alpha=0.
\end{equation}
I.e., the entropy $s(\epsilon, n)$ determines the relationship between
conserved densities $\epsilon$, $n$ and the corresponding
thermodynamic quantities $T$ and $\mu$.

The quadratic form of second derivatives of the entropy must be
negative to ensure thermodynamic stability. This
form can be diagonalized using variables $m\equiv s/n$ (specific entropy)
and $p$ (pressure). The probability of small fluctuations in terms of
these variables is given by
\begin{equation}
  \label{eq:Pdmdp}
  \mathcal P \sim \exp\left\{ -
  \frac V2 \left(
    \frac{n^2}{c_p}(\delta m)^2
    +   \frac{\beta}{wc_s^2}(\delta p)^2 + \dots
    \right)\right\}
\end{equation}

Specific heat $c_p=Tn(\partial m/\partial T)_{p}$ diverges at the
critical point. This corresponds to the probability of fluctuations
developing a ``flat direction'' along which $\delta p=0$, where the
fluctuations of the specific entropy
$V\langle(\delta m)^2\rangle=c_p/n^2$ become large.  \footnote{Other
  thermodynamic quantities also develop large fluctuations at the
  critical point. For example, $V\langle(\delta n)^2\rangle=\chi_T T$,
  where $\chi_T=(\partial n/\partial\mu)_T$ is also divergent.  The
  specific entropy $m=s/n$ is special from the hydrodynamic point of
  view. Unlike, e.g., baryon density, $m=s/n$ is a normal hydrodynamic
  mode in ideal hydrodynamics, i.e., being a ratio of conserved densities
  specific entropy is a diffusive mode which
  does not mix with propagating sound modes.}
The non-monotonic behavior of fluctuations as the critical point is
being approached and passed during the QCD phase diagram scan has been
proposed as a signature of the QCD critical point \cite{Stephanov:1998dy,Stephanov:1999zu}.

Before proceeding, let us pause to note that in these lectures we
focus on the fluctuations intrinsic in any system which affords local
thermodynamic, statistical description.  These fluctuations are
determined by the equation of state, as discussed above.  In the
context of heavy-ion collisions we must distinguish these fluctuations
from the fluctuations which are determined by the initial state of the
system. For example, from the fluctuations of the initial geometry of
the system. In experiments, such separation is not always
trivial. Typically it involves selecting collisions with similar
centrality, i.e., similar collision geometry. The effects of the
initial fluctuations are also qualitatively different from those of
thermodynamic fluctuations in that the correlations induced by initial
fluctuations are longer range (in longitudinal rapidity space) than
the thermodynamic fluctuations we discuss \cite{Kapusta:2011gt}. Most
importantly, the $\snn$ dependence of the initial fluctuations does
not reflect the {\em non-monotonicity\/} inherent in the thermodynamic
fluctuations in the vicinity of the critical point.

\subsection{Universality and non-Gaussianity of critical  fluctuations}

Since the equation of state is universal near critical points, the
fluctuation phenomena are also universal. In this section we shall
describe some universal properties of the fluctuations which are
relevant for the search for the QCD critical point in heavy-ion collisions.

Since the coefficient of the $(\delta m)^2$ term in
Eq.~(\ref{eq:Pdmdp}) vanishes at the critical point, non-Gaussian terms
in Eq.~(\ref{eq:Pdmdp}), such as $\delta m^3$, $\delta m^4$ etc.,
become important. This
makes non-Gaussianity of fluctuations a telltale signature of the
critical point \cite{Stephanov:2008qz}.

Similarly, the vanishing of the
coefficient of $\delta m^2$ in Eq.~(\ref{eq:Pdmdp}) makes gradient
terms, such as $(\bm \nabla \delta m)^2$,
important and leads to the divergence of the correlation length $\xi$ of the
fluctuations at the critical point. The non-Gaussianity
of fluctuations and the divergence of the correlation length go hand
in hand since the limit
$V/\xi^3\to\infty$, which ensures the Gaussianity of the fluctuations
via central limit theorem, is in obvious
conflict with $\xi\to\infty$.

The fluctuations at length scales of order the correlation length
cannot be described by the variable $\delta m$, or by the uniform
``mean field'' $\delta m$, alone. Instead, the fluctuating spatially varying
field $\delta m(x)$ must be considered. The corresponding scalar field
theory --- $\phi^4$ theory in three dimensions, is universal in that it
describes a wide range of critical phenomena from liquid-gas critical
points to uniaxial (Ising) ferromagnets.  The full details of the
universal theory of critical phenomena are beyond the scope of these
lectures, and are covered in many textbooks and reviews on critical
phenomena \cite{Amit:1984ms,Pelissetto:2000ek}. Here we shall emphasize only the
basic properties which are most relevant for the critical point search
in the beam-energy scan heavy-ion collision experiments.

The universality means that we can consider fluctuations of $m$, or
any other quantity, whose fluctuations diverge at the critical point,
such as baryon number, or entropy density (and in the case of ferromagnets,
magnetization density) as an order parameter and map it onto the field
variable $\phi$ in the $\phi^4$ theory, i.e.,
\begin{equation}
\Delta m\equiv m-m_{\rm c} \sim \phi,\label{eq:m-phi}
\end{equation}
where $m_{\rm c}$ is the value of $m$ at the critical point and the
implicit coefficient of proportionality is determined by the normalization of
the field $\phi$. The universality of the critical point phenomena
means that the probability of fluctuations can be expressed in terms of the
field $\phi$:
\begin{equation}
  \label{eq:P-phi}
  \mathcal P[\phi] \sim \exp\left\{
    -V\left( -h \phi + \frac r2 \phi^2 + \frac {\lambda}{4} \phi^4
      + \frac 12 (\bm\nabla\phi)^2
      \dots 
  \right)\right\}
\end{equation}
where we expanded in powers of $\phi$ and kept the leading
non-Gaussian term $\phi^4$ as well as the leading gradient term
$(\bm\nabla\phi)^2$.
At the critical point the
ordering field $h$ and the reduced temperature $r$ vanish. The
probability becomes non-Gaussian and, at the same time, the
correlation length $\xi$, proportional to $r^{-1/2}$ in the mean-field
approximation (i.e., neglecting fluctuations), diverges.
The cumulants of the order parameter $\phi$ also diverge,
according to
\begin{equation}
  \label{eq:phi-xi}
  \langle\phi^2\rangle \sim \xi^{2.}\,,
  \quad
   \langle\phi^3\rangle \sim \xi^{4.5}\,,
  \quad
   \langle\phi^4\rangle^{\rm c} \sim \xi^{7.}\,,
\end{equation}
with (approximate) powers different from mean-field values due to fluctuations.

\begin{figure}[h]
  \centering
  \includegraphics[height=12em]{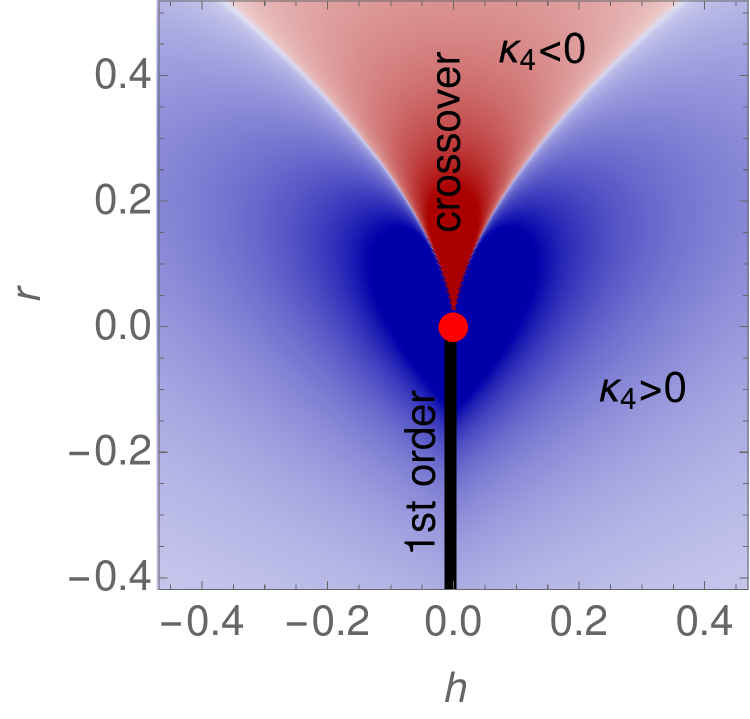}
  \hskip 3 em
\raisebox{2em}{\includegraphics[height=10em]{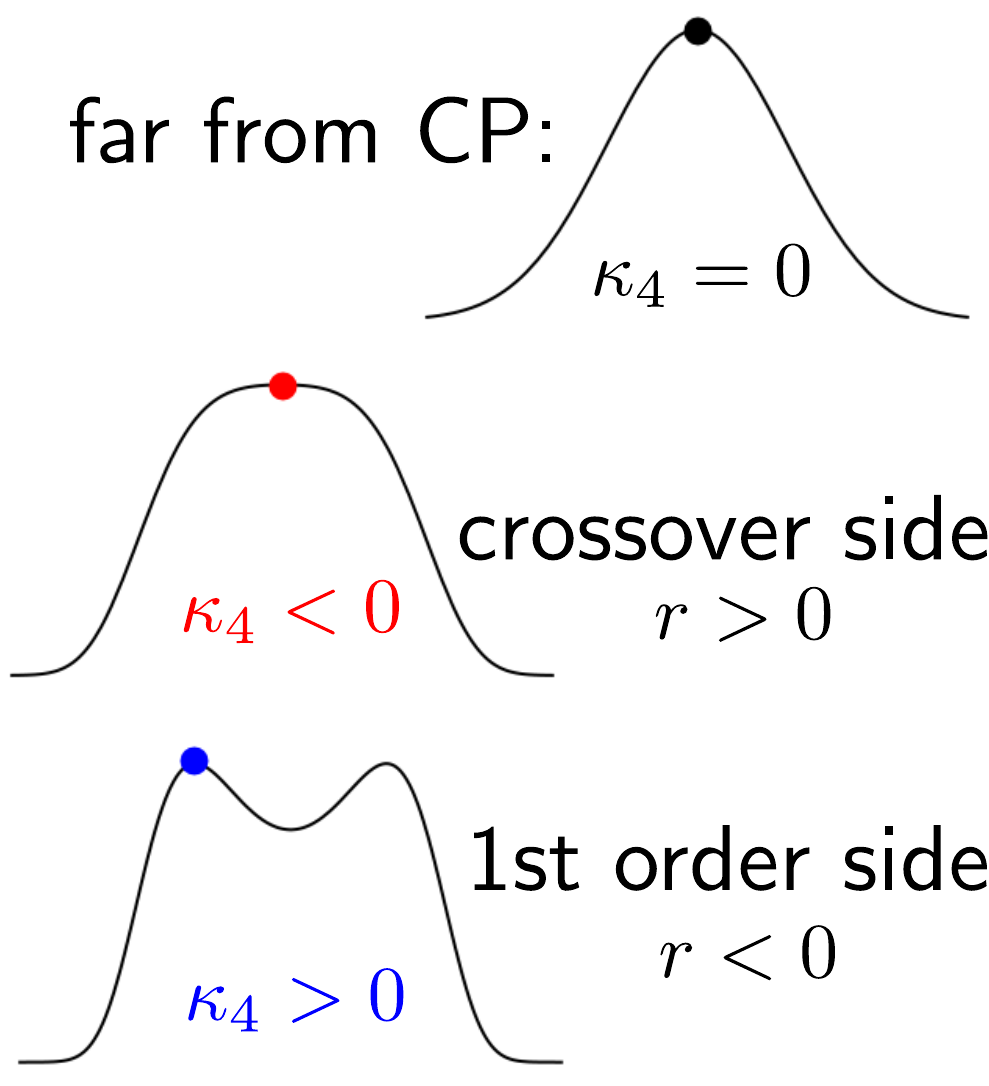}}
\caption[]{{\em Left:} The sign of the quartic cumulant
  $\kappa_4\equiv\langle\phi^4\rangle^c$ as a function of $\phi^4$
   (Ising model) field theory parameters $h$ and $r$. {\em Right:}
   Schematic representation of the probability distribution $\mathcal P[\phi]$
   in Eq.~(\ref{eq:P-phi}) at different points on the $h=0$ line.  }
\label{fig:Ising-rh}
\end{figure}

It is important that, unlike the variance $\langle\phi^2\rangle$,
which measures the width of the probability distribution, the cubic
and quartic cumulants, which measure the shape, are not positive
definite. The cubic cumulant, measuring the skewness of the
probability distribution $\mathcal P$, has the sign determined by the
sign of $h$. The sign of the quartic cumulant can be understood by
looking at the value of the cumulant along the crossover line $r>0$ at
$h=0$, as illustrated in Fig.~\ref{fig:Ising-rh}. The distribution
along this line starts off as Gaussian away from the critical point
and then splits into two maxima on the phase coexistence (first-order
phase transition) line for $r<0$. As $r\to+0$ the distribution becomes
``flatter'' which is represented by the negative value of the quartic
cumulant. \footnote{The quartic cumulant is related to kurtosis, $K$:
  $K=\langle\phi^4\rangle/\langle\phi^2\rangle^2$, which has an
  advantage of canceling overall normalization of $\phi$.  }

For $r<0$, on the coexistence line, while the distribution is
symmetric, the symmetry is spontaneously broken and the fluctuations
occur around one of the maxima. This skewness effect dominates
the quartic cumulant making it positive (see, e.g.,
\cite{Stephanov:2011pb} for more details).

The linear mapping from the Ising model variables $h$ and $r$ and the QCD
variables $\mu$ and $T$ captures the singular behavior of the QCD
thermodynamics due to the fluctuations of the order parameter field
$\phi$, as well as due to the next-to-leading relevant operator
($\phi^2$ in the mean-field approximation) \cite{Rehr:1973zz,Nicoll:1981zz,Pradeep:2022eil}.

Such a mapping has been standardized in Ref.~\cite{Parotto:2018pwx} using 6
parameters: $T_c$, $\mu_{Bc}$, $w$, $\rho$, $\alpha_1$ and
$\alpha_2$. The parameters $T_c$ and $\mu_{Bc}$ set the location of
the QCD critical point, while $\alpha_1$ is the angle of the slope of
the coexistence line (first-order phase transition line) at the
critical point in the $T$ vs $\mu_{B}$ plane. It is also the slope of
$m=m_{\rm c}$ (critical isentrope)
line at the critical point and is obtained by mapping
the zero magnetization line $\phi=0$ (i.e., zero magnetic field $h=0$) of the
Ising model onto the QCD phase diagram. The angle $\alpha_2$ is the
angle of the line on the QCD phase diagram onto which the constant
temperature line passing through the Ising critical point maps.

The non-Gaussian cumulants of the order parameter such as $m$ are
proportional to those in the Ising model (or $\phi^4$ theory,
see Eq.~(\ref{eq:m-phi})),
but mapped into the QCD phase diagram. In particular, these cumulants
contain singular contributions which diverge at the critical point
with universal powers of the correlation length, given approximately
by~\cite{Stephanov:2008qz}
\begin{equation}
  \label{eq:cumulants-xi}
  \Delta \langle(\delta m)^2 \rangle\sim \xi^{2.},\quad
  \Delta \langle(\delta m)^3 \rangle\sim \xi^{4.5},\quad
  \Delta \langle(\delta m)^4 \rangle^c \sim \xi^{7.}
\end{equation}
where $\Delta$ reminds us that this is a {\em contribution} to cumulants,
singular at the critical point. There is, of course, less singular and
regular contributions, or baseline.
The search for the critical point is aimed at detecting the
non-monotonic dependence of fluctuations measures on the collision
energy $\snn$ as the critical point is approached and passed, i.e.,
as the correlation length increases and then shrinks back to
non-critical background, or baseline, values.

\begin{figure}[h]
  \centering
  \includegraphics[height=12em]{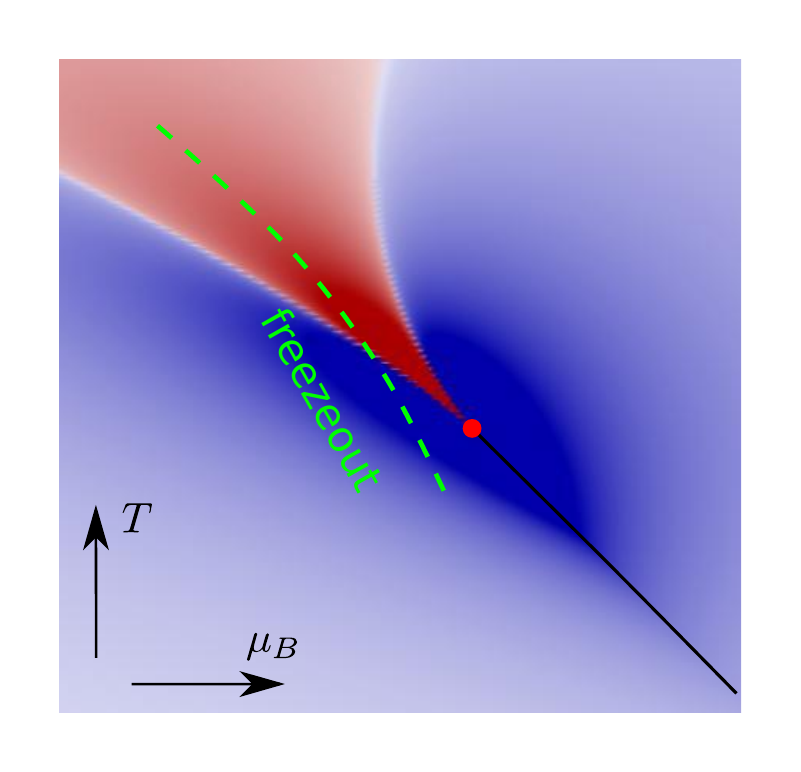}
  \hskip 2em
    \includegraphics[height=12em]{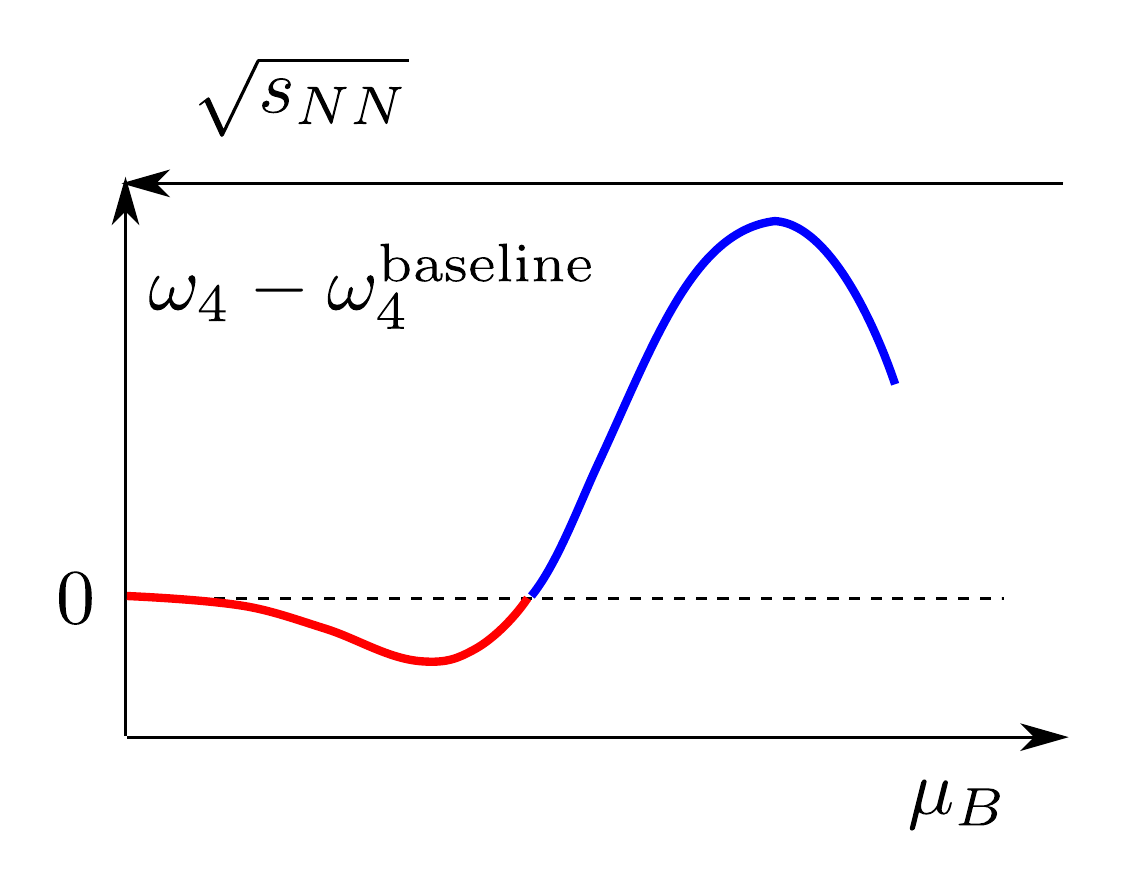}\\
    (a)\hskip 15em (b)
    \caption[]{The equilibrium expectation for the quartic cumulant of
      fluctuations as a function of temperature and baryon chemical
      potential on the QCD phase diagram in the vicinity of the
      critical point. Red and blue colors reflect the sign of the
      cumulant -- negative and positive respectively. Compare to Fig.~\ref{fig:Ising-rh}. The sign changes
      as the QCD phase diagram is scanned by varying $\snn$, and
      thus $T$ and $\mu_B$, along the freeze-out ``trajectory'' (dashed
      green line).}
    \label{fig:Tmu-kappa4}
\end{figure}

For example, the quartic cumulant around the QCD critical
point is illustrated in
Fig.~\ref{fig:Tmu-kappa4}(a).~\cite{Stephanov:2011pb}
The resulting
dependence on the collision energy, as it is varied in the region where
the freeze-out occurs near the critical point, is illustrated in
Fig.~\ref{fig:Tmu-kappa4}(b). The characteristic non-monotonicity of
this cumulant is one of the signatures of the critical point searched
for in the Beam Energy scan experiments \cite{Bzdak:2019pkr}.

The most recently published experimental measurements of
quartic cumulants by STAR collaboration indicate a non-monotonic
dependence of the type similar to the one shown in
Fig.~\ref{fig:Tmu-kappa4}(b) (see, e.g., Fig.4(2) in Ref.\cite{STAR:2020tga}). While the magnitude of the cumulant
sensitively depends on the EOS, and thus hard to predict (hence the
lack of the vertical scale on Fig.~\ref{fig:Tmu-kappa4}(b)) the
experimental results indicate that monotonic dependence is excluded at
3.1$\sigma$ level \cite{STAR:2020tga,STAR:2021iop,STAR:2021fge}.
While not yet a definitive indication of the
presence of the QCD critical point, this intriguing result motivates
the second phase of the beam energy scan (BES-II)
program which has collected higher
statistics data being processed and analyzed currently.

We shall focus on the recent
progress achieved in connecting these equilibrium fluctuations to the
experimental measurements. There are at least two steps that need to
be made
to get from the equilibrium thermodynamic fluctuations to the
experimental observables:

First, like thermodynamic variables themselves, the fluctuations of
these variables evolve. This evolution is subject to conservation laws
which may lead to significant memory effects, i.e., the lag of the
fluctuations behind the instantaneous equilibrium values.

Second, the thermodynamic/hydrodynamic quantities are not measured by
experiments directly. An important step is connecting the fluctuations
of these quantities to the observables,
such as measures of particle multiplicity fluctuations and correlations.

The following two sections will review the recent development towards
accomplishing these two steps necessary for connecting the theory and
the experiment.

\section{Fluctuations in hydrodynamics}

\subsection{Stochastic hydrodynamics}

There has been significant recent progress in understanding and
describing the {\em dynamics} of local thermodynamic fluctuations in
hot {\em relativistic} fluids, such as the QCD matter making up a heavy-ion
collision fireball.

The evolution of the QCD matter in the heavy-ion fireball is
successfully described in the framework of hydrodynamics. This
description is inherently statistical, and therefore, fluctuations are
an essential part of this description. One can also understand this
fact by invoking fluctuation-dissipation theorem according to which
dissipative systems (such as non-ideal hydrodynamics) have to be
stochastic. 

There are two complementary approaches for describing dynamics of
fluctuations, which we shall refer to as stochastic and
deterministic. Both begin with the Landau-Lifshits theory of
hydrodynamic fluctuations~\cite{Landau:2013stat2}. In this theory,
generalized to relativistic context in Ref.\cite{Kapusta:2011gt} , the
equations of hydrodynamics are {\em stochastic}. 

Hydrodynamic description exists because of the separation of
scales. The timescale of relaxation to local equilibrium (typical
microscopic time, such as inter-collision time) is much shorter that
the time needed to transport conserved quantities and establish
global equilibrium. This is typically a diffusion time proportional to
the square of the size of inhomogeneities in the system. This latter
slow evolution process is essentially governed by conservation
equations,
which we can write generically as
\begin{equation}
  \label{eq:dJJ}
  \partial_\mu \breve {\mathcal J}^\mu =0
\end{equation}
by combining all conserved currents into an array
\begin{equation}
  \label{eq:JJ==}
\breve{\mathcal J}^\mu \equiv \{\breve N_q^\mu, \breve T^{\mu\nu}\}\,.
\end{equation}
We use `breve' accent to denote stochastic (fluctuating) quantities.
For QCD such
conserved currents include currents of energy and momentum,
$T^{\mu\nu}$, as well as conserved charge currents, $N_q^\mu$, among
which we shall focus on the baryon charge $q=B$.

While each current involves one time and three spatial components, there is
only one equation for each current in (\ref{eq:dJJ}).  Therefore, to
close the system of equations we need to express all 4 current
components in terms of 1 variable -- the conserved density. To
maintain relativistic covariance, we want this density to be a
covariant object (scalar for the baryon charge density, or vector for
energy-momentum density). The density in the lab frame would not
do.

The fluid itself, however, allows us to define (in each space-time
point) a reference frame associated
with fluid's motion. This frame is often referred as the ``local rest frame''
of the fluid. Following Landau, we chose for that purpose the frame in
which the momentum density vanishes. The 4-velocity of such a frame
solves the following equation (known as ``Landau condition''):
\begin{equation}
  \label{eq:Teu}
  \breve T^\mu_{~\nu}\breve u^\nu = \breve\epsilon\breve u^\mu.
\end{equation}
Since $\breve T^\mu_{~\nu}$ fluctuates, so does $\breve u^\mu$.
Eq.~(\ref{eq:Teu}) also defines the energy density $\breve\epsilon$ in
the fluid's rest frame. The rest frame charge density is defined
similarly $\breve n_q=\breve u_\mu \breve N_q^\mu$.

It is convenient
to define an array of these covariant
variables -- one for each conservation equation in (\ref{eq:dJJ}):
\begin{equation}
  \label{eq:psi=uJ}
  \breve\Psi \equiv\breve u_\mu\breve{\mathcal J}^\mu = \{\breve n_q,\breve\epsilon\breve u^\nu\}\,.
\end{equation}
In order to close the system of equations we need to express all
components of the currents in $\breve{\mathcal J}^\mu$ in terms of the covariant
variables in $\breve\Psi$. The separation of scales in hydrodynamics
means that this
relationship is {\em local}. I.e., the currents $\breve{\mathcal J}(x)$
are functions of variables $\Psi(x)$ and its gradients at the same
point $x$:
\begin{equation}\label{eq:Jpsi=}
  \mathcal J^\mu[\Psi] =
  \{ nu^\mu, \epsilon u^\mu u^\nu - p(g^{\mu\nu}-u^\mu u^\nu) \}
  + \mbox{diffusive/viscous gradients}\,.
\end{equation}

The key point of the Landau-Lifshits theory of hydrodynamic
fluctuations is that the constitutive relations such as
(\ref{eq:Jpsi=}) are only satisfied {\em on average} -- hence the
absence of `breve' in Eq.~(\ref{eq:Jpsi=}). For fluctuating quantities there is random
discrepancy, which is due to the microscopic degrees of freedom
excluded from hydrodynamic description. Hence,
\begin{equation}\label{eq:bJpsi=}
  \breve{\mathcal J}^\mu =
  {\mathcal J}^\mu[\breve \Psi] + \breve{\mathcal I}^\mu\,,
\end{equation}
where $\breve{\mathcal I}^\mu$ is the local noise, $\langle\breve{\mathcal I}^\mu\rangle=0$, whose correlation function,
\begin{equation}
  \label{eq:II-delta}
  \langle\breve{\mathcal I}^\mu(x_1)\breve{\mathcal I}^\nu(x_2) \rangle
  \sim \delta^{(4)}(x_1-x_2)\,,
\end{equation}
is determined by
the fluctuation dissipation theorem, ensuring that the equilibrium
fluctuations and correlations have the correct magnitudes in agreement
with thermodynamics.

The evolution of fluctuations can be then described by, for example,
directly solving this system of stochastic equations
(\ref{eq:dJJ}),~(\ref{eq:bJpsi=}). In a numerical simulation, this
{\em stochastic} approach, however, produces a problem (also known as
infinite noise problem) due to the fact that the noise is singular at
$x_1=x_2$ in Eq.~(\ref{eq:II-delta}). The resulting solutions are
dependent on the hydrodynamic cutoff, i.e., the finite elementary
hydrodynamic cell size, $b$, complicating the ``continuum limit''
$b\to0$. Some solutions to this problem within a numerical simulation
have been proposed and implemented in the literature
\cite{Singh:2018dpk,Chattopadhyay:2023jfm}, but we shall not discuss
them here.

\subsection{Deterministic approach to hydrodynamic fluctuations}
\label{sec:determ-appr-hydr}

The approach which deals with the infinite noise problem {\em before}
the actual numerical simulation is performed has been also developed
recently for Bjorken
flow\cite{Akamatsu:2017,Akamatsu:2018,Martinez:2018}, for arbitrary
relativistic flow\cite{Stephanov:2017ghc,An:2019osr,An:2019csj}, and for non-Gaussian fluctuations\cite{An:2020vri,An:2022jgc}.
In this {\em deterministic} approach one expands in fluctuations around the
average $\Psi\equiv\langle\breve\Psi\rangle$:
\begin{equation}
  \label{eq:deltaPsi}
  \breve\Psi = \Psi + \delta\Psi\,,
\end{equation}
thus obtaining stochastic equations for the
fluctuations $\delta\Psi$ on the deterministically
evolving inhomogeneous fluctuation averaged background $\Psi(x)$.
These equations can then be used to derive {\em deterministic}
equations obeyed by the {\em correlation functions} of the fluctuations,
i.e., by averages of the products of the fluctuations, such as
$\langle\delta\Psi(x_1)\delta\Psi(x_2)\rangle$. 

The short-distance singularity of the noise results in ultraviolet
divergences in the deterministic equations -- the infinite noise
problem. Indeed expanding constitutive equations (\ref{eq:bJpsi=}) to
quadratic order and averaging one finds:
\begin{equation}
  \label{eq:<J>}
  \langle \breve{\mathcal J}^{\mu}(x)\rangle =
  {\mathcal J}^{\mu}[\Psi(x)]
  + \frac12\frac{\partial^2\mathcal
    J^\mu}{\partial\Psi_a\partial\Psi_b}
  \langle\delta\Psi_a(x)\delta\Psi_b(x)\rangle+\ldots
\end{equation}
where $a,b$ index hydrodynamic variables in the array
(\ref{eq:psi=uJ}).  The last term in Eq.~(\ref{eq:<J>}) is singular
because the correlator of hydrodynamic variables is evaluated at
coinciding points, and in (and near) equilibrium
$\langle\delta\Psi_a(x_1)\delta\Psi_b(x_2)\rangle\sim\delta(x_1-x_2)$.
The support
of the delta function is of the size of the hydrodynamic cell $b$,
over which the operators defining conserved densities are averaged to
obtain classical (stochastic) hydrodynamic variables $\breve\Psi$.

The key observation is that these singular (divergent in the limit $b\to0$)
contributions are {\em local} due to the hydrodynamic separation of
scales we discussed already. Thus the divergent terms in
Eq.~(\ref{eq:<J>}) must  simply
renormalize the local terms already present in the ``bare''
constitutive relations in Eq.~(\ref{eq:Jpsi=}).
The resulting system of renormalized
equations for hydrodynamic evolution of the renormalized average
densities, as well as the correlation functions, is ultraviolet finite,
i.e., cutoff independent, and the continuum limit can be taken (see,
Refs.\cite{An:2019osr,An:2019csj} for more detail). 

After the renormalization, the averaged hydrodynamic equations keep the
form of usual hydrodynamic equations to first order in gradients with
renormalized, i.e., physical, equation of state and transport
coefficients. However,  finite (i.e., cutoff independent)
fluctuation contributions appear beyond that order. These terms
introduce contributions non-local in space (effectively being of order
3/2 in gradients) and also non-local in time, leading to the phenomena
known as ``long-time tails''. Near the critical point these
fluctuation effects also lead to critical slowing down of the
relaxation to equilibrium \cite{Stephanov:2017ghc} and the related 
divergence of kinetic coefficients, most
notably, of bulk
viscosity~\cite{onuki2002phase,Stephanov:2017ghc,An:2019csj}.

\subsection{Multipoint Wigner transform}

Because fluctuations occur on shorter length scales than the
hydrodynamic evolution of the
background~\cite{Akamatsu:2017,Stephanov:2017ghc,An:2019osr} it is
convenient to consider
Wigner-transformed equal-time correlation functions. For a fluid at
rest globally, the Wigner transform definition is straightforward:
\begin{equation}
  \label{eq:WAB}
  W_{ab}(t,\bm x;\bm q) = \int {d^3\bm y}
  \left\langle\delta\Psi_a\left(t,\bm x + \frac{\bm y}{2}\right)
    \delta\Psi_b\left(t,\bm x - \frac{\bm y}{2}\right) \right\rangle
  e^{-i\bm q\cdot\bm y}\,,
\end{equation}
where $\delta\Psi_{a}$ is the fluctuation of a hydrodynamic
field from array (\ref{eq:psi=uJ})  labeled by index~$a$.

Non-Gaussianity of fluctuations, important for the critical point
search, is described by {\em connected} correlation functions of $k>2$
fluctuation fields:
\begin{equation}
  \label{eq:Hphi}
  H_{\listk{a}}(t,\clistk{\bm x})\equiv \left\langle
      \delta\Psi_{a_1}(t,\bm x_1)\dots\delta\Psi_{a_k}(t,\bm x_k)\dots
  \right\rangle^\textrm{connected}
\end{equation}
The corresponding generalization of Wigner transform was introduced in
Ref.\cite{An:2020vri} in terms of the Fourier integral with fixed
midpoint
\begin{equation}
  \label{eq:1}
  \bm
x\equiv \frac{\bm x_1 +\dots+ \bm x_k}{k}\,,
\end{equation}
i.e.,
\begin{multline}
  \label{eq:WAB...}
  W_{\listk{a}}(t,\clistk{\bm q})\equiv
  \int {d^3\bm y_1}\, e^{-i\bm q_1\cdot\bm y_1}
  \dots
    \int {d^3\bm y_k}\, e^{-i\bm q_k\cdot\bm y_k}
    \\
  \times\delta^{(3)}\left(\frac{\bm y_1+\dots+\bm y_k}{k}\right)
  H_{ab\dots}(t,\clistk{\bm x+\bm y})
\end{multline}
Due to the delta-function factor in
Eq.~(\ref{eq:WAB...}), the function $W$ does not change if all $\bm q$'s
are shifted by the same vector. This means that one of the $\bm q$
arguments is redundant. In practice, it is sufficient to know the
function $W$ at all values of $\bm q$ which add up to zero. In
particular, the correlator $H_{\listk{a}}$ can be obtained via inverse
transformation:
\begin{multline}
  \label{eq:HAB...WAB}
  H_{\listk{a}}(t,\clistk{\bm x}) =
    \int \frac{d^3\bm q_1}{(2\pi)^3}\,  e^{i\bm q_1\cdot\bm y_1}
    \dots
    \int \frac{d^3\bm q_k}{(2\pi)^3}\, e^{i\bm q_k\cdot\bm y_k}
  \\
  \times(2\pi)^3\delta^{(3)}\left(\bm q_1+\dots+\bm q_k\right)
  W_{\listk{a}}(t,\bm x;\listk{\bm q})\,.
\end{multline}
where $\bm y_i\equiv \bm x_i - \bm x$.
For $k=2$ the generalized Wigner function
$W_{ab}(t,-\bm q,\bm q)$ coincides with the usual 2-point Wigner function
defined in Eq.~(\ref{eq:WAB}).

\subsection{Evolution equations for fluctuations in a diffusion problem}

The hierarchy of evolution equations was
derived in Ref.\cite{An:2020vri} for fluctuations of density in a diffusion
problem, where the only hydrodynamic variable is the diffusing charge
density $n$ which obeys conservation equation and the Fick's law with
local noise
\begin{equation}
  \label{eq:dtn}
  \partial_t n = - \bm\nabla\cdot\bm N, \quad
  \bm N = -\lambda(n)\bm\nabla\alpha(n) + \mbox{noise}\,.
\end{equation}

\begin{figure}[h]
  \centering
  \includegraphics[height=9em]{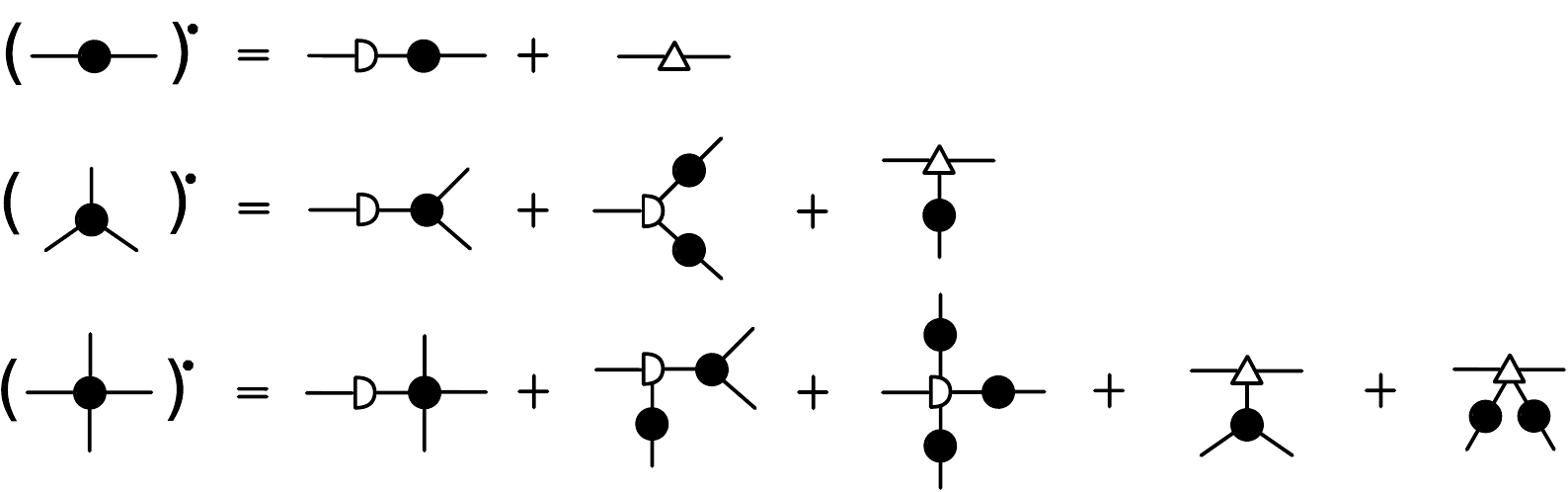}
  \caption{Diagrammatic representation of
    Eqs.~(\ref{eq:W_234}). See Refs.\cite{An:2020vri,An:2022jgc}. }
  \label{fig:diagrams}
\end{figure}

The evolution equations describe the relaxation of the
$k$-point functions $W_k$ of fluctuations of density $n$ to
equilibrium values given by thermodynamics in terms of the equation of
state $\alpha(n)$ (chemical potential divided by temperature):
\begin{subequations}\label{eq:W_234}
  \begin{equation}\label{eq:W_234-a}
    \partial_tW_2(\bm q)
    =-2\left[\gamma\bm{q}^2W_2(\bm{q})-\lambda\bm{q}^2
       \right]
       \,,
     \end{equation}
     \begin{multline}\label{eq:W_234-b}
       \partial_tW_3(\bm q_1,\bm{q}_2,\bm{q}_3)
       =-3\left[\gamma\bm{q}_1^2W_3(,\bm{q}_2,\bm{q}_3)
         +\gamma'\bm{q}_1^2W_2(\bm{q}_2)W_2(\bm{q}_3)
         \right.\\\left.
         +2\lambda'\bm{q}_1\cdot\bm{q}_2W_2(\bm{q}_3)\right]_\qpermiii
       \,,
     \end{multline}
     \begin{multline}\label{eq:W_234-c}
       \partial_tW_{4}(\bm{q}_1,\bm{q}_2,\bm{q}_3,\bm q_4)
       =-4\left[
         \gamma\bm{q}_1^2W_{4}(,\bm{q}_2,\bm{q}_3,\bm{q}_4)
         \right.\\\left.
           +3\gamma'\bm{q}_1^2W_2(\bm{q}_2)
           W_3(,\bm{q}_3,\bm{q}_4)
     +\gamma''\bm{q}_1^2W_2(\bm{q}_2)W_2(\bm{q}_3)W_2(\bm{q}_4)
        \right.
    \\
    \left.
      +3\lambda'\bm{q}_1\cdot\bm{q}_2W_3(,\bm{q}_3,\bm{q}_4)
      +3\lambda''\bm{q}_1\cdot\bm{q}_2W_2(\bm{q}_3)W_2(\bm{q}_4)
    \right]_\qpermiv
       ,
     \end{multline}
\end{subequations}
 where $\gamma=\lambda\alpha'$ is the diffusion coefficient. In
 Eqs.~(\ref{eq:W_234}) we suppressed arguments $t$ and $\bm x$, as
 they are the same for all functions $W_k$. Furthermore, note that all
 arguments of each function $W_k$ add up to zero -- reminiscent of the
 momentum conservation in Feynman diagrams. Therefore, to save space,
 we omitted the first argument in $W_k$ where this argument can be
 inferred from the condition $\bm q_1+\dots+\bm q_k=0$. For example,
 $ W_2(\bm q)\equiv W_2(,\bm q)\equiv W_2(-\bm q,\bm q)$ or
 $W_3(,\bm q_3,\bm q_4)\equiv W_3(-\bm q_3-\bm q_4,\bm q_3,\bm
 q_4)$. The symbol $ [\dots]_{\overline{1\dots k}} $ denotes the sum
 over all permutations of $\bm q_1$,\dots,$\bm q_k$\ divided by
 $1/k!$, i.e., the average over all permutations. Diagrammatic
 representation of Eqs.~(\ref{eq:W_234}) is given in
 Ref.\cite{An:2020vri} (see also Fig.~\ref{fig:diagrams}).

Equations~(\ref{eq:W_234}) represent the leading order in hydrodynamic
gradient expansion. The next order would correspond to loop diagrams
(in terms of the diagrammatic technique used in
Fig.~\ref{fig:diagrams}). It represents the feedback of fluctuations.
For two-point functions, such feedback have been considered in
Refs.\cite{Akamatsu:2017,Akamatsu:2018,Martinez:2018,Stephanov:2017ghc,An:2019osr,An:2019csj}
and the corresponding renormalization of hydrodynamic equations has
been shown to remove cutoff dependence due to ``infinite noise''
problem, as discussed above in Section~\ref{sec:determ-appr-hydr}.

\subsection{Confluent formalism for arbitrary relativistic flow}

For a {\em relativistic\/} fluid with nontrivial velocity gradients
the definition of the equal-time correlator in Eq.~(\ref{eq:Hphi}) and
its Wigner transform are insufficient since the rest frame of the
fluid is different at different space-time points, and the concept of
``equal time'' is thus ambiguous. To maintain Lorentz covariance in
the description of fluctuations, Ref.\cite{An:2019osr}, and for
non-Gaussian fluctuations, Ref.~\cite{An:2022jgc},
propose to use averaged local rest frame of the fluid at the midpoint
$x\equiv(t,\bm x)$ of the correlator to define ``equal time''. The
same locally defined frame is also used to define, or measure,
fluctuations of velocity.

The idea is to equip every space-time point $x$ with an orthonormal
triad of basis 4-vectors $e_{\mathring{a}}(x)$, $\mathring a=1,2,3$,
or $\bm e(x)$, spatial in the local rest frame of the fluid at that
point, i.e.,
\begin{equation}
\bm e(x)\cdot u(x)=\bm 0.\label{eq:e-dot-u}
\end{equation}
One can then define equal-time $k$-point correlators
as functions of $k$ space-time points expressed as
\begin{equation}
  x_1=x+\bm e(x)\bm\cdot\bm y_1,
\quad\ldots\quad,\quad
x_k=x+\bm e(x)\bm\cdot \bm y_k,\label{eq:x1...xk}
\end{equation}
in terms of the midpoint $x$ and the
separation 3-vectors $\bm y_1,\dots, \bm y_k$, which sum to zero: $\bm
y_1+\ldots+\bm y_k=0$.

Some of the fields in the correlator may not be Lorentz scalars, for
example, fluctuations of velocity, $\delta u$. One would
like to express such fluctuations in a frame associated with the fluid,
such as the local rest frame, rather than in an arbitrary lab frame. The
local rest frame of the fluid is, however, different for different
points $x_1,\dots,x_k$. To use the same frame for all points while
making sure that fluctuations represent deviations from fluid at rest
locally, we boost the fluctuation variables (if they are not scalars)
from the rest frame at the point they occur, say, at $x+y_1$, to that
at the midpoint $x$ of the correlator. This operation leads to the
{\em confluent correlator} defined as:
\begin{equation}
  \label{eq:confluent-correlator-def}
   H_{a_1\dots a_k}(x_1,\dots,x_k) \equiv
   \left\langle
     \left[\Lambda(x,x_1)\delta\Psi(x_1)\right]_{a_1}\dots
     \left[\Lambda(x,x_k)\delta\Psi(x_k)\right]_{a_k}
    \right\rangle^{\rm c}
\end{equation}
where $\Lambda(x,x_1)$ performs the boost on the corresponding
fluctuation field $\delta\Psi(x_1)$ such that the 4-velocities in points
$x_1$ and $x$ are related by
\begin{equation}
  \label{eq:Lambda-u}
  \Lambda(x,x_1) u(x_1) = u(x).
\end{equation}
Superscript `c' means ``connected'' as in
Eq.~(\ref{eq:Hphi}).

We can now take the multipoint Wigner transform of the confluent
correlator we just defined in Eq.~(\ref{eq:confluent-correlator-def})
with respect to the 3-vectors
$\bm y_1,\dots,\bm y_k$ describing
separation of the points in the local basis at point $x$
(cf. Eq.~(\ref{eq:WAB...})):
\begin{multline}
  \label{eq:W...H}
  W_{a_1\dots a_k}(x;\bm q_1,\dots,\bm q_k)\equiv
    \int {d^3\bm y_1}\, e^{-i\bm q_1\bm\cdot\bm y_1}\dots
    \int {d^3\bm y_k}\, e^{-i\bm q_k\bm\cdot\bm y_k}
    \\
  \times\delta^{(3)}\left(\frac{\bm y_1+\dots+\bm y_k}{k}\right)
   H_{a_1\dots a_k}(x+\bm e(x)\bm\cdot\bm y_1,\dots,x+\bm e(x)\bm \cdot \bm y_k)
 \end{multline}
The inverse is given by (cf. Eq.~(\ref{eq:HAB...WAB}))
\begin{multline}
  \label{eq:H...W}
  H_{a_1\dots a_k}(x+\bm e_1(x)\bm\cdot\bm y_1,\dots,
  x+ \bm e_k(x)\bm\cdot\bm y_k) \\ =
    \int \frac{d^3\bm q_1}{(2\pi)^3}\,  e^{i\bm q_1\bm\cdot\bm y_1}\dots
    \int \frac{d^3\bm q_k}{(2\pi)^3}\, e^{i\bm q_k\bm\cdot\bm y_k}
    \\\times  
    \delta^{(3)}\left(\frac{\bm q_1+\dots+\bm q_k}{2\pi}\right)
  W_{a_1\dots a_k}(x;\bm q_1,\dots,\bm q_k)\,.
\end{multline}

Care must
be taken also in defining a derivative  with respect to the midpoint
of such a correlation
function in order to maintain the ``equal-time in local rest frame''
condition. The local rest frame is different in points $x$ and
$x+\Delta x$ used to define the derivative.
A derivative which maintains ``equal-time'' condition is introduced in
Ref. \cite{An:2019osr,An:2019csj} and termed
{\em confluent derivative\/}.
It is defined via the $\Delta x\to0$ limit of the following equation:
\begin{multline}
  \label{eq:confluent-nabla-W}
  \Delta x\cdot\bar\nabla  W_{a_1\dots a_k}(x;\bm q_1,\dots,\bm q_k)
  \equiv
   \Lambda(x,x+\Delta x)_{a_1}^{b_1}\dots
   \Lambda(x,x+\Delta x)_{a_k}^{b_k}
   \\\times  W_{b_1\dots b_k}(x;\bm q_1',\dots,\bm q_k')
   - W_{a_1\dots a_k}(x;\bm q_1,\dots,\bm q_k)\,,
\end{multline}
where
\begin{equation}
  \label{eq:q'q}
  \bm
q'=\bm e(x+\Delta x)\cdot[\Lambda(x+\Delta x,x)\bm e(x)\bm\cdot\bm
q]\equiv R(x+\Delta x,x) \bm q\,.
\end{equation}
In Eq.~(\ref{eq:confluent-nabla-W}) the (non-scalar) fluctuation
fields are boosted from point $x+\Delta x$ back to point $x$,
similarly to the definition of the confluent correlator in
Eq.~(\ref{eq:confluent-correlator-def}). In addition, at $x+\Delta x$
we evaluate the function using the set of 3-wavevectors $\bm q_i'$
given by Eq.~(\ref{eq:q'q}), different from the set $\bm q_i$ used at
point $x$. The new set is obtained by representing each vector $\bm q$
as a {\em 4-vector\/} orthogonal to $u(x)$, $\bm e(x)\cdot\bm q$, then
boosting this vector to the rest frame at point $x+\Delta x$ and then
expressing it again as a 3-vector, but now in the basis
$\bm e(x+\Delta x)$ orthogonal to $u(x+\Delta x)$. The resulting
transformation of vector $\bm q$ is a rotation, denoted by $R$ in
Eq.~(\ref{eq:q'q}).

Taking the limit $\Delta x\to0$ in Eq.~(\ref{eq:confluent-nabla-W})
one can express confluent derivative as follows:
\begin{equation}
  \label{eq:nabla-W}
  \bar\nabla_\mu W_{a_1\dots a_2} = \partial_\mu W _{a_1\dots a_2}
  + k\left(
    \mathring\omega^{\mathring a}_{\mu\mathring b}q_{1\mathring a}
    \frac{\partial}{\partial q_{1\mathring b}}W_{a_1\dots a_k}
    - \bar\omega^{b}_{\mu a_1}W_{ba_2\dots a_k}
    \right)_{\overline{1\dots k}}\,,
\end{equation}
where the confluent connection $\bar\omega$ is a generator of the
infinitesimal boost~$\Lambda$ and $\mathring\omega$ is a generator of
the infinitesimal rotation $R$:
\begin{align}
  \label{eq:Lambda-omega}
  \Lambda(x+\Delta x,x)^a_{~b} = \delta^a_b - \Delta
  x^\mu\bar\omega^a_{\mu b},
  \\ \label{eq:R-omega}
  R(x+\Delta x,x)^{\mathring a}_{~\mathring b}
  = \delta^{\mathring a}_{\mathring b}
  - \Delta x^\mu\mathring\omega^{\mathring a}_{\mu\mathring b}\,.
\end{align}

The indices $a,b,a_1\dots a_k$ label
fluctuating fields. The confluent connection $\bar\omega^a_{\mu
b}$ is nonzero when indices $a,b$ refer to different components of a
Lorentz vector (such as $\delta u$). In this case the connection
satisfies
\begin{equation}
  \label{eq:nabla-u}
  \bar\nabla_\mu u^\alpha \equiv \partial_\mu u^\alpha
  + \bar\omega^\alpha_{\mu\beta}u^\beta=0,
\end{equation}
(local velocity is ``confluently'' constant), which follows from
Eqs.~(\ref{eq:Lambda-u}) and~(\ref{eq:Lambda-omega}).%
\footnote{For a scalar field (e.g., energy
density, pressure fluctuations, etc) the confluent connection is, of course,
zero. E.g., $\bar\nabla_\mu \delta m = \partial_\mu \delta m$.}

The rotation connection $\mathring\omega$ is determined by
Eqs.~(\ref{eq:q'q}) and~(\ref{eq:R-omega}):
\begin{align}
  \mathring\omega^{\mathring b}_{\mu\mathring a}
  =
  e^{\mathring b}_\alpha\left(\partial_\mu e^\alpha_{\mathring a}
  + \bar\omega^\alpha_{\mu\beta}e^\beta_{\mathring a}\right)\,.
\end{align}
Naturally, it satisfies:
\begin{equation}
  \label{eq:nabla-e}
  \bar\nabla_\mu e^\alpha_{\mathring a} \equiv \partial_\mu
  e^\alpha_{\mathring a}
  + \bar\omega^\alpha_{\mu\beta}e^\beta_{\mathring a}
  - \mathring\omega^{\mathring b}_{\mu\mathring a}e^\alpha_{\mathring b}=0\,
\end{equation}
(local basis vectors are confluently constant).

The boost $\Lambda$ is not defined uniquely by Eq.~(\ref{eq:Lambda-u})
-- only up to a rotation keeping $u$ unchanged.  The simplest choice
is the boost without additional rotation, which corresponds to
confluent connection given explicitly by
\begin{equation}
  \label{eq:omega-udu}
  \bar\omega^\alpha_{\mu\beta}=u_\beta\partial_\mu u^\alpha
  - u^\alpha\partial_\mu u_\beta\,.
\end{equation}
For this choice of the confluent connection the rotation connection also
simplifies to
\begin{align}
  \mathring\omega^{\mathring b}_{\mu\mathring a}
  =
  e^{\mathring b}_\alpha
  \partial_\mu e^\alpha_{\mathring a}\,.
\end{align}

\subsection{Evolution equations for hydrodynamic fluctuations}

There are five normal hydrodynamics modes, which can be described as two
propagating modes and three diffusive.\footnote{We focus on
  hydrodynamics involving baryon charge. Each additional charge adds one diffusive mode to the count.} The propagating modes
correspond to fluctuations of pressure mixed with the fluctuations of
the longitudinal (with respect to the wave vector $\bm q$)
velocity. The frequency of these modes is $c_s|\bm q|$. The diffusive
modes are the fluctuations of the specific entropy $m\equiv s/n$ at
fixed pressure and
transverse velocity. The relaxation rate of these modes is
proportional to the square of their wavenumber $\bm q^2$. The
slowest diffusive mode near the critical point is specific entropy
because its diffusion constant 
vanishes at the critical point.

In this review we shall focus on the slowest diffusive mode $m$ for two
reasons. First, because it is the slowest and therefore the furthest from
equilibrium mode. Second, in equilibrium this mode shows the
fluctuations of the order parameter (i.e., $\delta m\sim\delta\phi$, Eq.~(\ref{eq:m-phi})), divergent at the critical point.

The evolution equation for the Wigner-transformed confluent two-point
correlator of the specific entropy fluctuations, $\langle\delta
m\delta m\rangle$, derived in
Ref.\cite{An:2019csj} reads:
\begin{equation}
  \label{eq:LW}
  \mathcal L[W_{mm}(\bm q)]
  = (\partial\cdot u) W_{mm}(\bm q)
  - 2\gamma_{mm}\bm q^2\left[W_{mm}(\bm q)
  - \frac{c_p}{n^2}\right]\,.
\end{equation}
where $\gamma_{mm}\equiv\kappa/c_p$ -- the heat diffusion constant,
and $\mathcal L$ is
the Liouville operator given by
\begin{equation}
  \label{eq:mathcalL}
  \mathcal L [W_{mm}] \equiv \left[
    u\cdot\bar\nabla
    - \partial_\nu u^\mu (\bm e_\mu\cdot\bm q) \left (\bm e^\nu
    \cdot \frac{\partial}{\partial \bm q}\right)\right]W_{mm}\,.
\end{equation}
The first term is the confluent derivative along the flow of the
fluid, while the second describes stretching and/or rotation of the
vectors $\bm q$ due to the expansion and/or rotation of the fluid. In
the case of expansion one can think of this term as describing the
Hubble-like ``red shift'' of the fluctuation wavevector $\bm q$.
For example, for Bjorken flow the Liouville operator takes the form
\begin{equation}
  \label{eq:L-Bj}
  \mathcal L_{\rm Bj} = \partial_\tau
  - \frac{q_3}{\tau}\frac{\partial}{\partial q_3},
\end{equation}
where the second term describes the ``stretching'' of the
fluctuations, i.e., the ``red shift'' of the wave number $q_3$ due to
longitudinal expansion.\footnote{Naturally, we have chosen the triad
  of the 4 vectors $\bm e$ in such a way that the spatial part of the
  4-vector $e_3$ points along the direction of the longitudinal flow,
  while the $e_1$ and $e_2$ are constant.  In this case the rotation
  connection ($\mathring\omega$) terms vanish in the confluent
  derivative in Eq.~(\ref{eq:nabla-W}). The confluent connection
  ($\bar\omega$) terms are absent already for arbitrary flow because
  the fluctuating quantity, $m$, is a scalar. }

The first term on the r.h.s. of Eq.~(\ref{eq:LW}), describes the
scaling of the fluctuation magnitude with the volume of a hydrodynamic
cell, as the cell expands at the rate $\partial\cdot u$. This trivial rescaling
could be absorbed by multiplying $W$ by a conserved density, such as
the baryon density, $n$. The equation for a rescaled function $N_{mm}\equiv
nW_{mm}$ is the same as in Eq.~(\ref{eq:LW}), but without the
$\partial\cdot u$ term.
\begin{equation}
  \label{eq:LN}
  \mathcal L[N_{mm}(\bm q)]
  = 
  - 2\gamma_{mm}\bm q^2\left[N_{mm}(\bm q)
  - \frac{c_p}{n}\right]\,.
\end{equation}

The last term in  Eq.~(\ref{eq:LW}) describes diffusive relaxation of
fluctuations towards equilibrium given by thermodynamic quantity
$c_p/n^2$ (or $c_p/n$ for $N_{mm}$).

It is instructive to compare Eq.~(\ref{eq:LW}) for $W_{mm}$ (or the
corresponding Eq.~(\ref{eq:LN}) for $N_{mm}$) to
Eq.~(\ref{eq:W_234-a}) for the density-density correlator
$\langle\delta n\delta n\rangle$ in the diffusion problem. The main
difference is that the time derivative is replaced by the Liouville
operator, which takes into account the flow of the fluid. The
$(\partial\cdot u)$ term on the r.h.s. (absent when Eq.~(\ref{eq:LW})
is written in terms of $N_{mm}$, Eq.~(\ref{eq:LN})) is also an effect of
the flow -- expansion. The diffusive relaxation terms are different
because the correlated quantities are different, $\delta m\delta m$ in
Eq.~(\ref{eq:LW}) and $\delta n\delta n$ in Eq.~(\ref{eq:W_234-a}). The
coefficients, however, can be mapped onto each other via substitution
\begin{equation}
  \label{eq:nmsubstitution}
  n\to m,\quad \gamma=\lambda\alpha'\to\gamma_{mm}=\frac{\kappa}{c_p},
  \quad \lambda \to \frac{\kappa}{n},
  \quad W_2\to N_{mm}.  
\end{equation}

The evolution equations for non-Gaussian correlators $W_{mmm}$ and
$W_{mmmm}$ (or $N_{mmm}\equiv n^2W_{mmm}$ and $N_{mmmm}\equiv n^3W_{mmmm}$)
are qualitatively similar to those in the diffusion problem
in Eq.~(\ref{eq:W_234-b}) and~(\ref{eq:W_234-c}). However, unlike the
case of $W_{mm}$, where the whole equation~(\ref{eq:LN}) can be
obtained from Eq.~(\ref{eq:W_234-a}) for $W_2$ by the substitution
given by Eq.~(\ref{eq:nmsubstitution}), only the terms containing
leading singularities at the critical point can be obtained from
Eq.~(\ref{eq:W_234-b}) and~(\ref{eq:W_234-c}) by the
substitution~(\ref{eq:nmsubstitution}) (with $W_3\to N_{mmm}$ and
$W_4\to N_{mmmm}$). There are subleading
singularities, which are due to the nonlinearity of the mapping $n\to
m$. These terms are written explicitly in
Ref.\cite{An:2022jgc}.

\section{Freeze-out of fluctuations and observables}

The previous section was devoted to recent progress in describing the
evolution of fluctuations in hydrodynamics using correlators of
hydrodynamic variables in {\em coordinate} space. Heavy-ion collision
experiments do not measure such densities, or their correlations,
directly. Instead the particle multiplicities and their correlations
in {\em momentum} space are measured. In this section we describe how
to connect the theoretical description in terms of fluctuating
hydrodynamics to these experimental observables.

\subsection{Event-by-event fluctuations  and
  their experimental measures}\label{sec:event-event-fluct}

Typical experimental measures are cumulants of the event-by-event
fluctuations or correlations of particle multiplicities.
For example, if $N_p$ is the proton number in an event, its fluctuation
in the event is $\delta N_p\equiv N_p - \langle N_p\rangle$ and
$\langle(\delta N)^2\rangle$ is its quadratic cumulant, or variance, where
$\langle\dots\rangle$ is the event average. Higher-order cumulants,
measuring non-Gaussianity of fluctuations, are constructed similarly
(see, e.g., Ref.\cite{Bzdak:2019pkr} for a review).

In addition, correlations between particles can be also measured, such
as $\langle\delta N_p\delta N_\pi\rangle$ -- a correlation between
proton and pion multiplicities. Such measures can also include
higher-order correlators\cite{Athanasiou:2010kw}.
Similarly to correlations between species one can also consider
correlations between particles with different momenta.

The number of particles $d^3N_i$ in a given momentum cell $d^3p$ is
given by the well-know Cooper-Frye formula~\cite{Cooper:1974mv}
\begin{equation}
  \label{eq:Ni-f}
  d^3N_i = \frac{d^3p}{(2\pi)^3} \frac{p^\mu}{p^0}\int_\Sigma d^3\Sigma_\mu(x) f_i(x,p)
\end{equation}
in terms of the phase space distribution
function $f_i(x,p)$ integrated over the freeze-out
hypersurface $\Sigma$. Therefore the correlations between different momentum
cells and/or between different species, can be expressed in terms of
the correlation functions of fluctuations of $f_i(x,p)$:
$\delta f\equiv f-\langle f\rangle$.

To simplify and shorten notations we shall combine the species index (which
includes all discrete quantum numbers, such as mass, spin, isospin,
etc.) together with the coordinate and momentum
into a single composite index $A=\{i,x,p\}$.
Therefore, the general correlator which, upon integration over the
freeze-out hypersurface, gives the observable correlation measures has the
form
\begin{equation}
  \label{eq:dfdf}
  \langle \delta f_{i_1}(x_1,p_1) \dots \delta f_{i_k}(x_k,p_k)\rangle \equiv
  \langle \delta f_{A_1} \dots \delta f_{A_k}\rangle\,.
\end{equation}

Usually the Cooper-Frye prescription (\ref{eq:Ni-f}) is applied to
determine the event-by-event {\em averaged} number of particles in
terms of the averaged distribution function $\langle f_A\rangle$. This
averaged function is expressed in terms of the hydrodynamic variables,
or fields, $T(x)$ and $\mu(x)$:
\begin{equation}
  \label{eq:f-T-mu}
  \langle f_A \rangle\equiv \langle f_{i_A}(x_A,p_A)\rangle=\left[
    \exp\left\{\beta(x) u(x)\cdot p_A - \alpha(x) q_A
    \right\} - (-1)^{2s_A}
  \right]^{-1}
\end{equation}
where $\beta=1/T$, $\alpha=\mu/T$, $q_A$ is the charge of the particle $A$
with respect to the chemical potential $\mu$ (baryon charge, for
example) and $s_A$ is the spin of the particle.

In order to convert
hydrodynamic {\em fluctuations} into particle event-by-event
fluctuations we need an analogous freeze-out prescription for {\em
  correlators} in Eq.~(\ref{eq:dfdf}).

\subsection{Freeze-out of fluctuations and the maximum entropy method}

The generalization of the Cooper-Frye freeze-out to fluctuations has
been first considered in Ref.\cite{Kapusta:2011gt}.  In the approach
of Ref.\cite{Kapusta:2011gt} the fluctuations of the phase space
distribution function $f(x,p)$ are assumed to be caused by
fluctuations of the hydrodynamic variables/fields $T(x)$ and $\mu(x)$
on which $f(x,p)$ depends, as in Eq.~(\ref{eq:f-T-mu}). As a result,
coordinate space correlations in $T(x)$ and $\mu(x)$ translate into
the phase space correlations in $f(x,p)$.

This approach has an important flaw, which becomes obvious if
one considers fluctuations in an (almost) ideal gas. In this case,
there are fluctuations of hydrodynamic variables, such as charge
density $n(x)$, but there are no momentum space correlations of
$f(x,p)$, which would, nevertheless, be produced if the approach of
Ref.\cite{Kapusta:2011gt} were to be applied. Instead, the hydrodynamic
fluctuations of $n(x)$ are matched on the particle side by trivial
(Poisson, in the ideal gas case) {\em uncorrelated} fluctuations of the
occupation numbers in each phase-space point.

This problem has been addressed in
Refs.\cite{Ling:2013ksb,Kapusta:2017hfi} by subtracting this trivial
ideal gas contribution from hydrodynamic fluctuations of $n(x)$ before
applying the procedure of Ref.\cite{Kapusta:2011gt} to the remainder,
which is due to interactions and out-of-equilibrium dynamics.

Generalization of this approach to fluctuations of other hydrodynamic
variables, and to non-Gaussian fluctuations, proved
elusive until the principle of maximum entropy for fluctuations was
proposed and implemented in Ref.\cite{Pradeep:2022eil}.
In this approach the matching of conserved hydrodynamic densities such as
$\breve\epsilon(x)$ and $\breve n_q(x)$, defined in
Eqs.~(\ref{eq:Teu}),~(\ref{eq:psi=uJ}),
is done  on
an event-by-event basis. For example, we have to match the fluctuations
of conserved charge density:
\begin{equation}
  \label{eq:dndf}
  \delta n_q (x) = \sum_i \int_{p} q_i \delta f_i(x, p) \equiv
  \int_{\tilde A}q_A \delta f_A(x) \equiv
  \int_A q_A \delta^{(3)} (x-x_A) \delta f_{A}\,,
\end{equation}
where $\int_p$ is a 3-integral over momenta with the Lorentz-invariant
measure and $\sum_i$ is a sum over the species of particles with id
label $i$ (corresponding to mass, spin, isospin, etc.) carrying charge
$q_i$ corresponding to density $n_q$ (e.g., baryon charge when
$n_q=n_B$ is the baryon density). We have also introduced a convenient
shorthand $\int_{\tilde A}$, which denotes the sum and the momentum
space integral together (but no space integration).
One can think of the composite index $A$ as
describing not only the particle ``id'' (i.e., mass, spin, etc) but
also its position in the phase space (as if each particle species
has its own phase space): $A\equiv\{i_A,p_A,x_A\}$.
Finally, we also introduced $\int_A$, which includes integration over
the whole phase space (momentum $p_A$ {\em and} coordinate $x_A$) of each
particle species, and a convenient shorthand
$f_A\equiv f_A(x_A)$. The delta function simply reflects the locality of
freeze-out (i.e., each hydrodynamic cell at point $x$
is converted into particles located at the same position $x_A=x$).
Similarly, matching of the energy-momentum density requires
\begin{equation}
  \label{eq:dedf}
  \delta ( \epsilon (x) u^\mu(x)) =
  \int_{\tilde A}p^\mu \delta f_A(x)\equiv\int_A p^\mu \delta^{(3)} (x-x_A) \delta f_{A}\,.
\end{equation}
It is convenient to organize equations such as (\ref{eq:dndf})
and~(\ref{eq:dedf}) into an indexed array, where lowercase index $a$
runs through five hydrodynamic variables
$\delta\Psi_{a}=\delta\{n_q,\epsilon u^\mu\}$, similar to
Eq.~(\ref{eq:psi=uJ}):
\begin{equation}
  \label{eq:dpsidf}
  \delta\Psi_a\equiv\delta\Psi_{a}(x_a) = \int_A P^A_a \delta f_A\,,
\end{equation}
where
\begin{equation}\label{eq:PAa}
  P^A_a=\{q_A,p_A^\mu\}\delta^{(3)}(x_a-x_A)
\end{equation}
is the array of the contributions of a single particle at point $x_A$ to
hydrodynamic densities $\Psi_{a}=\{n,\epsilon u^\mu\}$ in a
cell around point
$x_a$ on the freeze-out surface. Similarly to particle index $A$ it is
convenient to view hydrodynamic field index $a$ as a composite index
labeling both the field and the point on the freeze-out surface where
the value of this field is measured.

Eq.~(\ref{eq:dpsidf}) for fluctuations imply relationships
between the hydrodynamic correlators in space points
$x_{a_1}\dots x_{a_k}$,
\begin{equation}
  \label{eq:Hcorr}
  H_{\listk{a}}\equiv \langle\delta\Psi_{a_1}\dots\delta\Psi_{a_k}\rangle\,,
\end{equation}
and particle correlators in phase-space points $A_1\dots A_k$,
\begin{equation}
  \label{eq:Gcorr}
  G_{\listk{A}}\equiv \langle\delta f_{A_1}\dots\delta f_{A_k}\rangle\,,
\end{equation}
which have the form:
\begin{equation}
  \label{eq:HG}
  H_{\listk{a}} = \int_{\listk{A}} G_{\listk{A}}P^{A_1}_{a_1}\dots P^{A_k}_{a_k}\,.
\end{equation}

Equations (\ref{eq:HG}) represent constraints on the particle
correlators $G_{AB\dots}$ imposed by conservation laws. These
constraints alone are not enough to completely determine
$G_{AB\dots}$ simply because there are more ``unknowns'' $G_{AB\dots}$
then the constraints.
The situation is similar already for ensemble
(i.e., event) averaged quantities, or one-point functions. In
this case the knowledge of
the averaged energy, momentum and baryon density  is not
sufficient alone to determine the particle distribution functions
$f_A$. Additional input is needed.

In the absence of additional
information, the most natural solution is the one which maximizes the
entropy of the resonance gas into which the hydrodynamically evolved
fireball freezes out. That entropy is given by the
functional of $f_A$:
\begin{equation}
  \label{eq:S_f}
  S[f_A]= \int_A(\theta_A^{-1}+f_A)\ln (1+\theta_Af_A)-f_A \ln f_A\,,
\end{equation}
where $\theta_A=(-1)^{2s_A}$ is determined by the particle spin
$s_A$. \footnote{Neglecting quantum statistics, i.e., taking
  $\theta_A\to0$ one obtains the familiar Boltzmann entropy
  $ S[f_A]=\int_A f_A(1-\ln f_A)\,.  $} Maximizing $S[f_A]$, subject to
the constraints on $f_A$ imposed by conservation laws, produces the
well-known equilibrium distribution $f_A$ given in
Eq.~(\ref{eq:f-T-mu}), underlying the Cooper-Frye freeze-out
procedure, which has been widely and successfully used for describing
experimental data for half a century. This observation have been also
used recently to describe systematically deviations from equilibrium
due to viscous or diffusive gradients in Ref.~\cite{Everett:2021ulz}.

In order to apply the maximum entropy approach to {\em fluctuation}
freeze-out, on needs the entropy of fluctuations as a functional of
$f_A$  as well as correlators $G_{AB}$, $G_{ABC}$, etc. Conceptually, this
entropy, $S[f_A,G_{AB},\dots]$
represents the (logarithm) of the number of the microscopic
states in the resonance gas ensemble with the given set of
correlators. The single-particle entropy $S[f_A]$ is the value of
$S[f_A,G_{AB},\dots]$ when all correlators $G_{AB\dots}$ are given by
their values in the equilibrium resonance gas.

The expression for the
functional $S[f_A,G_{AB},\dots]$ was found in
Ref.\cite{Pradeep:2022eil}. For example, keeping only terms with
out-of-equilibrium two-point correlators it reads \footnote{The calculation of the entropy of fluctuations along these lines for a
two-point correlator of {\em hydrodynamic} variables can be found in
Ref.~\cite{Stephanov:2017ghc}, and it is
{\em mathematically} similar to the 2-PI action in quantum field theory~\cite{Luttinger:1960ua,Baym:1962sx,Cornwall:1974vz,Norton:1974bm,Berges:2003pc}.}
\begin{equation}
  \label{eq:S2}
    S_2[f,G] = S[f]
  + \frac12{\rm Tr\,}\left[
\log(CG) - CG + 1
\right]\,.
\end{equation}
The last term is always negative except for $G=C^{-1}\equiv\bar G$ which is the
equilibrium value of the correlator $G$, where $C_{AB}=-\delta^2
S[f]/(\delta f_A\delta f_B)$. When $G=\bar G$ the last term vanishes
and $S_2$ is maximized with respect to~$G$.

However, maximizing the entropy in Eq.~(\ref{eq:S2}) with respect to $G$
under constraints in Eq.~(\ref{eq:HG}) gives
\begin{equation}
  (G^{-1})^{AB} = (\bar G^{-1})^{AB} + (H^{-1}-\bar H^{-1})^{ab}
\, P_{a}^{A}\, P_{b}^{B}\,.
\label{eq:G2}
\end{equation}
In this equation and below the repeated lower case indices $a$, $b$,
etc. imply summation over the set hydrodynamic variables
$\Psi_a$, $\Psi_b$ {\em and} volume integration over
hydrodynamic cells at points $x_a$, $x_b$. Due to the delta
functions in the definition of $P_a^A$ in Eq.~(\ref{eq:PAa}), these implied
integrals in Eq.~(\ref{eq:G2}) simply set the spatial arguments of
$(G^{-1})^{AB}$ to those of $(H^{-1})^{ab}$, i.e., $x_A=x_a$ and
$x_B=x_b$. When hydrodynamic correlator $H$ equals its equilibrium
value $\bar H$ in the resonance gas, the particle correlator $G$ equals
$\bar G$ --- its value in the resonance gas (i.e., $f_A\delta_{AB}$,
neglecting quantum statistics).

If deviations of fluctuations from equilibrium resonance gas are
small, the equation~(\ref{eq:G2}) can be linearized in such
deviations. The deviations could be due to non-equilibrium effects,
which have to be small for hydrodynamics to apply, or due to effects
of the critical point. Linearized equation relates deviations of the
particle correlators $\Delta G_{AB}= G_{AB}-\bar G_{AB}$ to the
deviations of the hydrodynamics correlators $\Delta H_{ab}=H_{ab}-\bar
H_{ab}$ from the resonance gas values:
\begin{equation}
  \label{eq:DG2}
  \Delta G_{AB}
  =
  \Delta H_{ab}
  (\bar H^{-1}P\bar G)^a_A (\bar H^{-1}P\bar G)^b_B,
\end{equation}

Similarly, the non-Gaussian cumulants $G_{ABC\dots}$ of particle
fluctuations can be expressed in terms of the non-Gaussian cumulants
of the hydrodynamic variables $H_{abc\dots}$. Such non-linear relations
similar to Eq.~(\ref{eq:G2})
can be derived from the corresponding entropy functional found in
Ref.\cite{Pradeep:2022eil} and we will not reproduce them
here.

Linearized relations valid for small deviations from the
equilibrium resonance gas, however, are simple and instructive.
The relationship
is similar to Eq.~(\ref{eq:DG2}), but instead of the ``raw'' deviations
from equilibrium $\Delta G_{AB\dots}$ and $\Delta H_{ab\dots}$, i.e.
correlations relative to equilibrium, the
proportionality relation holds between {\em irreducible}
  relative correlators $\hat\Delta G_{AB\dots}$
defined in Ref.\cite{Pradeep:2022eil}. An irreducible
correlator $\hat\Delta G$ is different from the ``raw'', or reducible,
relative correlator
$\Delta G$ by subtraction of correlations
involving only a smaller subset of the points $AB\dots$. The irreducible $\Delta H$ differs from $\Delta
H$ similarly. The resulting linear relation
generalizes Eq.~(\ref{eq:DG2}):
\begin{equation}
  \label{eq:DhatGDhatH2}
  \widehat\Delta G_{\listk{A}}
  =
  \widehat\Delta H_{\listk{a}}
  (\bar H^{-1}P\bar G)^{a_1}_{A_1} \dots (\bar H^{-1}P\bar G)^{a_k}_{A_k},
\end{equation}
where $\widehat\Delta G$ and $\widehat\Delta H$
denote {\em irreducible} relative correlators for particles and for
hydrodynamic variables, respectively. For two-point (Gaussian)
correlators $\hat\Delta G=\Delta G$ and $\hat\Delta H=\Delta H$ and Eq.~(\ref{eq:G2})
reproduces Eq.~(\ref{eq:G2}).

Eq.~(\ref{eq:DG2}) thus solves the problem of translating fluctuations
in hydrodynamics into correlations between particles at freeze-out, in
such a way as to obey the conservation laws on event-by-event basis
(by satisfying constraints in Eq.~(\ref{eq:HG})). As one can see, it
systematically eliminates spurious ``self-correlations'' discussed in
the beginning of this section not only for Gaussian, but also for
non-Gaussian cumulants.

Similarly to the way the maximum entropy approach reproduces and
generalizes Cooper-Frye prescription for event {\em averaged}
observables, the maximum entropy approach to fluctuations reproduces,
justifies, and generalizes prior approaches to freezing out
fluctuations, in particular, of critical fluctuations, as shown in
Ref.\cite{Pradeep:2022eil}. Such a prior approach involving fluctuating
background field $\sigma$ was introduced in
Ref.\cite{Stephanov:1999zu}, generalized to non-Gaussian fluctuations
in Refs.\cite{Stephanov:2011pb,Athanasiou:2010kw}, and then further
generalized to {\em non-equilibrium} fluctuations in
Ref.\cite{Pradeep:2022mkf}.

The $\sigma$-field approach\cite{Stephanov:1999zu,Stephanov:2011pb,Athanasiou:2010kw}, however, besides the knowledge of QCD EOS,
requires the knowledge of the properties of the field $\sigma$ such as
its correlation length as well as its coupling to observed particles. These
properties would depend on the nature of this field -- an {\it a
  priori} unknown
mixture of scalar fields such as chiral condensate, energy and
baryon number densities. It was also not clear how to deal with 
non-critical fluctuations or contributions of lower point correlations
to higher-point correlators. All these uncertainties are absent in
the maximum entropy approach. The correlations described by
Eq.~(\ref{eq:DhatGDhatH2}) are very similar to the correlations
induced by the $\sigma$ field given by a mixture of hydrodynamic
fields determined by the QCD EOS itself.

The maximum entropy approach thus provides a
direct connection between the fluctuations of the hydrodynamic
quantities and the observed particle multiplicities, with their
fluctuations. This connection is determined by the EOS of QCD in the resonance
gas phase where the freeze-out occurs. The effects of the critical
point and non-equilibrium are encoded in the non-trivial correlations
described quantitatively by Eq.~(\ref{eq:DhatGDhatH2}).


\section{Summary and conclusions}

These lecture notes describe recent theory developments aimed at
mapping QCD phase diagram and the search for the critical point in
heavy-ion collisions.

The existence and location of the QCD critical point is a
major unanswered question about the QCD phase diagram. Universality of
critical phenomena allows us to draw predictions which do not require
precise microscopic knowledge of QCD dynamics.

In particular,
singular behavior of the fluctuations at critical points should
manifest itself in non-monotonic dependence of the fluctuation
measures in heavy-ion collisions as the critical point is approached
and then passed in the course of the beam-energy scan. This dependence
is especially pronounced for non-Gaussian fluctuation measures.

The dynamical nature of the heavy-ion collision fireball requires
treatment of fluctuations {\em beyond equilibrium} thermodynamics. This
means not only that we need to be able to describe fluctuations of
hydrodynamic variables, but also that we need to be able to translate
those hydrodynamic fluctuations into the fluctuations and correlations
of particle multiplicities observable in experiments. The
understanding of how to do this in a way consistent with
hydrodynamics, in particular, with conservation laws, has only emerged
recently. And these recent developments are the focus of the lectures.

As is often the case with a developing field of research, these
lectures can only attempt to capture a snapshot of the current state
of the art.  Some questions still require more careful analyses, and
some tools, such as fully-fledged simulation of the heavy-ion
collision with fluctuations, still need to be developed before
comparison to experiment can become quantitative and
reliable. Naturally, much of the future development of the field will
be informed by the experimental data from the BES-II program at RHIC
as well as from experiments at planned future heavy-ion collision
facilities~\cite{LRPNS:2023}.

I would like to express my gratitude to the organizers of the 63rd
Cracow School of Theoretical Physics for their warm hospitality in
Zakopane. This work is supported by the U.S. Department of Energy,
Office of Science, Office of Nuclear Physics Award
No. DE-FG0201ER41195.

\newpage 
\bibliographystyle{JHEPmod}
\bibliography{refs,fluctuations}

\end{document}